\documentclass[3p,preprint,sort&compress]{elsarticle}
\usepackage{amsmath}
\usepackage{amssymb}
\usepackage{amsfonts}
\usepackage{epsfig}
\usepackage{graphicx}
\usepackage{multirow}
\usepackage{caption}
\usepackage{cite}
\usepackage[usenames,dvipsnames]{xcolor}
\usepackage{tikz}

\journal{arXiv}

\begin{document}

\begin{frontmatter}

\title{Nationalism, Immigration and the Dynamics of Language Evolution}

\author[PUC]{Andr\'e Barreira da Silva Rocha\corref{cor1}}
\ead{andre-rocha@puc-rio.br}
\cortext[cor1]{corresponding author}
\address[PUC]{Department of Industrial Engineering, Pontifical Catholic University of Rio de Janeiro,\\ Rua Marqu\^es de S\~ao Vicente, 225, G\'avea, CEP22451-900, Rio de Janeiro, RJ, Brazil.}

\begin{abstract}
I study the interplay between language competition and ideology struggle in a country where there is a native high-status language
and a low-status language spoken by immigrants. Language transition is governed by a three-state model similar to the Minett-Wang (2008) and the
Heinsalu et al. (2014) models. I introduce the novelty that, among natives, an ideological struggle exists between
a group of nationalists and a group of pro-immigrants, thus a two-state model as in Abrams-Strogatz (2003). When bilingualism
emerges as an equilibrium, there might be a completely segregated society with monolingual immigrants and monolingual nationalistic
natives, thus an undesired outcome contrasting with the widespread literature. Removing ideology struggle, bilingualism is never
stable, thus results in Heinsalu et al. (2014) are solely due to the way conversion rate constants are defined in their model. If language status
is explicitly defined in the model, bilingualism becomes doomed in the long run.
\end{abstract}


\end{frontmatter}

\section{Introduction}
\label{sec:intro}
The issues of opinion formation, population dynamics, ideology struggle, language competition and the dynamics of language death have attracted the interest of many physicists, particularly those applying the tools of statistical mechanics and complex systems. Despite the existence of a connection between those issues, studies dealing with their combined effects on society are still relatively rare. In this paper, I investigate the interplay between language competition and ideology struggle by formulating a Lotka-Volterra-type model in order to study the evolution of language and ideology in a country where there is a native high status language and a low status language spoken by immigrants coming from the same origin country. Both languages compete against each other for speakers as in \citep{Abrams03,Pinasco06,Mira05,Minett08,Patriarca14} and I introduce the novelty that, among natives, there is an ideological struggle between two groups of individuals, a group of nationalists and a group whose members welcome immigrants and might have an interest in their culture and language. Such kind of ideology struggle was studied in \citep{Barreira13} in a different framework using evolutionary game theory.

The relevance and motivation to integrate the evolution of ideology and language into the same model can be seen in \citep{Shafir95} who states that language and religion have been the two most important cultural infrastructures serving as bases for national differentiation and modern demands of autonomy in Europe. In the USA, a recent example of the interplay among nationalism, immigration and language took place in Dade County, Greater Miami, from the 1960s until the early 1990s and is described in the work of \citep{Garcia04}. After the Cuban Revolution, Cuban migration to the USA, which main destination was South Florida, took place initially in three waves, Jan/1959-Oct/1962, Sep/1965-Apr/1966 (freedom flights) and Apr-Oct/1980 (Mariel boatlift), respectively. The first two waves included skilled workers and high officials of the deposed government, which together with the Cold War, was an incentive for American natives to welcome those immigrants. As a result, in 1973, Dade County even declared itself officially bicultural. However, the third wave was composed of a large amount of working class immigrants, many of them black, triggering nationalism, anti-Cuban demonstrations and African American riots in South Florida. Those events culminated with voters repealing the bicultural status and Dade County declaring that it was unlawful the use of public funds to promote any language other than English or any culture other than that of the USA. 

The model developed here might be of interest in countries with a significant foreign population such as the case of Luxembourg, whose foreign population jumped from 26.3\% (1981) to 38.1\% (2003) of the total population. Portuguese nationals accounted for 13.7\% of the total population in Luxembourg in 2003 \citep{Luxembourg15}. In such a context, depending on the interaction between foreign individuals and natives (citizens), together with public policy regarding language status and assimilation, nationalism might find a fertile environment to develop and spread. The recent wave of mass immigration of refugees from Syria and other countries to Europe shows that a study of these issues is of utmost importance.

The contributions of this paper are twofold: (i) when nationalism is introduced in the model, results are in line with the literature in the sense that eventually one language might disappear with the entire population becoming monolingual either in the high or low status language. On the other hand, results also show that when bilingualism emerges, it might be related to a completely segregated society with monolingual immigrants and monolingual nationalistic natives. Hence, bilingualism might not be a desired outcome as widespread outlined in the literature so far; (ii) if ideology transition is removed, the model becomes a particular case of \citep{Minett08} with the extension of \citep{Patriarca14} with regard to the contribution of bilinguals to the transition probabilities. I show that under this framework, bilingualism is never stable, thus results in \citep{Patriarca14} are solely due to the way the four conversion rate constants $k_{ZX}$, $k_{XZ}$, $k_{ZY}$, $k_{YZ}$ (see next Section) are defined in their model, i.e., without any relation among them. When the constants are related to each other through language status, bilingualism becomes doomed and disappears in the long run. 

The remainder of the paper is organized as follows: in Section 2, I review some key papers in the literature on ideology struggle and language competition, discussing their main results and limitations. Section 3 presents the model and results discussing the interface between nationalism and language. Section 4 concludes. 

\section{Literature Review}
\label{sec:LR}
Models dealing with language competition attracted the interest of physicists particularly after the work of \citep{Abrams03}. The latter provides a minimal model for language shift, where two languages compete with each other for speakers. Individuals belong to one of two monolingual groups and the population has no spatial structure, i.e., individuals interact with others at the same rate, thus having each an equal probability in terms of language influence. Languages are assumed as fixed entities, i.e., grammar, syntax and other structural properties of language do not evolve over time. The probability of an individual shifting from one language to the other is proportional to the attracting language perceived status and its number of speakers, such that no one shifts to a language with no speakers or no status. Language status increases with the higher social and economic opportunities a typical speaker of a language can obtain. Apart from the strong assumption that no individual is bilingual, one drawback of the model is that both languages cannot coexist in equilibrium, which contrasts with the existence of bilingual societies in practice.

The absence of language coexistence was soon overcome in the literature. Although \citep{Pinasco06} also ignore the existence of bilingual speakers, they provide an alternative population dynamics setup to that of \citep{Abrams03}, belonging to the family of Lotka-Volterra models, in which the rate of change of the number of monolingual speakers of each competing language depends not only on a term related to the competition for speakers but also on an additional term taking into account the intrinsic growth rates of each population and their different carrying capacities in the absence of competition. Thus, when speakers from different languages do not interact, each language evolves according to uncoupled Verhulst equations towards their carrying capacities in the long run. Even assuming that only one language is attractive, when both languages compete for speakers, as long as a threshold condition is satisfied, the only stable equilibrium is one in which both languages coexist. The threshold condition requires: (i) a small and easily reachable carrying capacity of the population speaking the attractive language. The latter is achievable by increasing the opportunities afforded to the speakers of the low status language, (ii) a high growth rate of the speakers of the low status language and (iii) a low rate of language shift through minimizing contact between speakers of the two different languages.   

While language death seems to be a quite important issue and the model of \citep{Abrams03} displays a good fit with historical data for endangered languages such as Scottish Gaelic, Quechua and Welsh, which appear to be on track for extinction in the future, their model is unable to explain cases of successful bilingualism such as that of Spain with the survival of Catalan, Galician and Vasco, together with Castilian, even after the Francoist period during which only the latter was officially allowed. Apart from the attempt of \citep{Pinasco06} to explain such a puzzle, \citep{Mira05} also studied the issue of language coexistence and went one step further by introducing a group of bilingual speakers. Their work was particularly motivated by the language context in the region of Galicia (Northwest of Spain) where there exists a bilingual majority alongside Galician and Castilian monolingual minorities. Differently from \citep{Abrams03} where the endangered languages are very different from the high status language, Galician and Castilian are very similar and \citep{Mira05} extend the model of \citep{Abrams03} by incorporating a parameter to account for the similarity between both competing languages. Taking into account two populations of monolinguals and one of bilingual speakers, therefore modelling the evolution of the three groups over time through a system of two coupled differential equations, their model reduces to the model in \citep{Abrams03} if bilinguals do not exist and both languages are completely different. As in \citep{Abrams03}, the language transition probabilities are proportional to the attracting language status and its share of speakers in the total population, with bilinguals also contributing to the share. Transitions are possible between any group, i.e., $X\rightarrow Y$; $X\rightarrow Z$; $Y\rightarrow X$; $Y\rightarrow Z$; $Z\rightarrow X$; $Z\rightarrow Y$, where $X$, $Y$ and $Z$ represent respectively each of the two monolingual and the bilingual groups. Their main finding is that, given the status of both competing languages, if a minimum threshold for language similarity is satisfied, bilingualism becomes a stable equilibrium in which the bilingual group survives in the long run together with the group of monolingual individuals speaking the high-status language, thus the low-status language survives among the speakers of the former group. 

Hence, the model of \citep{Mira05} was successful in modelling both monolingual and bilingual speakers plus the survival of bilingualism while it could still reproduce the results in \citep{Abrams03}. On the other hand, two important criticisms were discussed in \citep{Minett08}. Firstly, when no monolingual speaker of one of the competing languages remains, there would be no advantage conferred to bilinguals and the latter should disappear in the long run. Also, regarding the possibility of direct language shift between $X$ and $Y$-speakers and vice-versa, such transitions would be very unlike in the real world as those would involve the simultaneous loss of one language and the acquisition of the competing one, i.e., children would acquire a different language from the one of their parents, in some cases making communication between them impossible. Or monolingual adults learning a new language would simultaneously forget their native language. In order to overcome those pertinent limitations, \citep{Minett08} extend the model in \citep{Abrams03} accounting for a bilingual group as in \citep{Mira05} and they introduce a distinction between horizontal and vertical transmission of language. Their mechanism of language transmission is more restrictive than that of \citep{Mira05} given they do not allow for direct transitions between monolingual groups. 
In terms of vertical transmission, they assume children of monolingual parents necessarily acquire the language of their parents as their first language while children of bilingual parents may acquire either or both of the competing languages, thus the following group transitions are allowed: $X\rightarrow X$; $Y\rightarrow Y$; $Z\rightarrow X$; $Z\rightarrow Y$ and $Z\rightarrow Z$. Regarding horizontal transmission, they assume bilingual adults remain bilingual throughout their lifetimes while monolingual adults may either remain monolingual or learn a second language. Thus, transitions $Z\rightarrow Z$; $X\rightarrow Z$; $Y\rightarrow Z$; $X\rightarrow X$ and $Y\rightarrow Y$ are allowed. Based on these assumptions, a three-state model similar to that in \citep{Mira05} is built, where the rate of change of the proportion of each of the three groups in the total population follows the vertical and horizontal language transmission mechanisms with probability $\mu$ (mortality rate, when adults are replaced by children) and $1-\mu$, respectively. Thus, transition from monolingualism to bilingualism necessarily takes place through horizontal transmission and transition in the opposite direction is made through vertical transmission. The system of ordinary differential equations (ODEs) modelling the evolution of monolinguals and bilinguals is then given by:
\begin{eqnarray*}
\dot{x}&=&\mu c_{ZX}s_{X}zx^a-(1-\mu)c_{XZ}s_Yxy^a \\
\dot{y}&=&\mu c_{ZY}s_{Y}zy^a-(1-\mu)c_{YZ}s_Xyx^a
\end{eqnarray*}
where $x\in X$, $y\in Y$, $z\in Z$, $s_i$ is the status of language $i$, $s_{-i}=1-s_i$ (as in \citep{Abrams03}), $x+y+z=1$, $\dot{z}=-\dot{x}-\dot{y}$, $c_{ij}$ is the conversion rate from group $i$ to group $j$ which can be associated with the level of educational resources available in a particular language or the rate of interaction between pairs of individuals belonging to groups $i$ and $j$ and $a$ is a parameter accounting for how the attractiveness of language $i$ scales with the share of its speakers. \citep{Abrams03} estimated $a$ to cluster about 1.31 while in \citep{Minett08} and \citep{Patriarca14} the value is assumed to be 1, i.e., the attractiveness of a language increases linearly in the proportion of its speakers. One difference between \citep{Minett08} and \citep{Mira05} is that in the former the transition probabilities are influenced only by the number of monolingual speakers of the language being acquired, thus bilinguals play no role on such probabilities. This follows an interpretation where, in addition to being able to communicate with both $i$ monolinguals and bilinguals, the $i$ monolingual adult who subsequently acquires language $j$ can additionally communicate with $j$ monolinguals, therefore only the proportion of monolingual speakers of $j$ matters. The system of ODEs has two stable equilibria, both associated with monolingualism and there are two critical unstable points, one related to pure bilingualism and another in which individuals from all three groups survive and co-exist. Thus, the base model of \citep{Minett08} shares the same prediction as in \citep{Abrams03}, i.e., one of the two competing languages will eventually become extinct. On the other hand, an additional stable fixed point can be created in the interior of the phase space by adopting a mechanism of intervention in which, when the proportion of speakers of the endangered language falls below a given threshold, the status of the endangered language is increased and the $c$-coefficients are adjusted such that the conversion rates from bilingualism to both monolingual groups are increased and the conversion rates regarding transitions in the other direction are decreased (i.e., the likelihood that all three groups survive increases by making the two languages become more isolated with incentives for monolingual education of children).

As can be seen, both models in \citep{Abrams03} and \citep{Minett08} require some sort of policy intervention for bilingual sustainability. A successful attempt to model language competition in which both the vertical and horizontal mechanisms of language transmission are respected and bilingualism can be sustainable without any form of intervention is proposed in \citep{Patriarca14}. They also present a three-state model but with distinct transition probabilities between groups than those of \citep{Minett08}. With such novel dynamics, bilingualism is asymptotically stable. They modify the model in \citep{Minett08} as follows:
\begin{eqnarray}
\dot{x}&=&k_{ZX}zx-k_{XZ}x(y+\alpha z)\label{P1} \\
\dot{y}&=&k_{ZY}zy-k_{YZ}y(x+ \beta z)\label{P2}
\end{eqnarray}
where the parameters $k_i$ combine effects such as mortality rate and language status. $\alpha,\beta\in\left[0,1 \right] $ represent respectively the importance of bilinguals as representatives of languages $Y$ and $X$ to the monolinguals of $X$ and $Y$. This is particularly important when the two competing languages have significant different statuses because in such a case the bilingual community would consist mainly of bilinguals coming from the low status language group and individuals from the high status language might not have an interest in learning the low status one
 (e.g.: if the $Y$-language has a very low status, $\alpha$ would be very low given that $Z$ would be mainly composed of $Y$-speakers who learned the high status $X$-language, thus bilingualism would be associated with the high importance of the $X$-language). In this framework, bilinguals play a role in the language transition probabilities as in \citep{Mira05} but in a more reasonable way. Note that parameter $a$ in the above system is set to $1$, giving a Lotka-Volterra-type model. The dynamics leads bilingualism to be asymptotically stable whenever $\alpha$ and $\beta$ are high enough such that they are simultaneously above critical values, i.e., the first necessary condition for stable bilingualism requires bilinguals to be sufficiently regarded by monolinguals of one language as members of the monolingual community of the competing language. The former condition for bilingualism can only be fulfilled if transitions from bilinguals to monolinguals are more rare than the opposite transitions, thus in contrast with \citep{Minett08}. Also, while in \citep{Patriarca14} bilingualism survives through bilinguals taking over the entire population, in \citep{Minett08} when bilingualism survives, all three groups coexist.

From the discussion so far, it can be seen that the seminal model of \citep{Abrams03} has been subsequently refined in the literature over time, with the incorporation of more realistic assumptions in an attempt to prove that bilingualism can be sustained in the long run, i.e., a low-status language is not necessarily doomed. Although some of the papers cited above were successful in this task, their policy recommendations display some clashing directions. Moreover, in the present work, I show that this sort of obsession with bilingualism might not be a positive issue in some contexts, particularly when there is an interplay between the issues of ideology struggle and language competition on society. 

Language competition and ideology struggle can be connected not only on the sociological
level but also in terms of modelling given that Lotka-Volterra-type models can be also
found in the literature predicting the evolution of competing ideologies over time. In \citep{Vitanov10} a model for ideological competition in a country with growing population is formulated. The evolution of the aggregate population follows the Verhulst law and individuals are divided in $n$ groups, each with a different specific ideology, plus one ideology-free group. The number of individuals belonging to each of these $n$ ideological groups evolves according to the following Lotka-Volterra equation:
\begin{equation*}
\dot{N}_i=r_iN_i+\sum_{j=0}^{n}f_{ij}N_j+\sum_{j=0}^{n}b_{ij}N_iN_j;\ i=1,\cdots,n
\end{equation*}
where the first term accounts for the death of followers of ideology $i$, i.e., $r_i<0$. The second term accounts for the so-called unitary conversion from ideology $j$ to ideology $i$, i.e., conversion motivated by the information environment such as newspapers and election speeches. $f_{ij}$ reflects the intensity of conversion and the latter is also proportional to the number of followers of ideology $j$, thus there is no direct contact between followers of different ideologies. The remaining term takes into account binary conversion, which is the one due to direct interaction between members of two ideologies, thus proportional to $N_iN_j$, with intensity $b_{ij}$.

Thus, a model unifying the impact of both ideology and language on society is presented in the next section. The background of language and ideology competition is based on \citep{Barreira13}. The latter uses a different type of model based on evolutionary game theory to study how nationalism and assimilation evolve among natives and immigrants, respectively, assuming that the latter group comes from the same origin country. In the host country, the population of natives is divided in two groups, one which members welcome immigrants and another which members display nationalistic behaviour and try to restrict the access of immigrants to public services. On the side of immigrants, there are also two groups: one which members display an interest to learn the native language of the host country, thus making an attempt to become culturally assimilated while the other group does not learn. When immigrants do not live in an enclave, thus being dispersed across the host country, depending on the parameters of the model, a dynamics of Lotka-Volterra type is obtained, with nationalism and immigrant assimilation oscillating over time. In contrast, in the present work, using a different framework, an asymptotically stable equilibrium is found.
 
\section{Model}
\label{sec:MD}
\begin{figure}[htp]
\centering
\begin{tikzpicture}%
  \draw[black,thick,fill=black!50] (-3,0) circle (0.75cm);
  \draw[black,thick,fill=black!50] (1.5,0) circle (0.75cm);
  \draw[black,thick,fill=black!50] (6,0) circle (0.75cm);
  \draw[black,thick,fill=black!50] (-3,-4) circle (0.75cm); 
  
  \node (a) at (-3,0)  { \huge{X} };
  \node (b) at (1.5,0)  { \huge{Z} };
  \node (c) at (6,0) { \huge{Y} };
  \node (d) at (-3,-4) { \huge{W} };
  
  \draw[line width=1.5pt,->] (-2.35,0.35) .. controls (-0.75,1.5) .. (0.85, 0.35);
  \node at (-0.75,1.5) {$(1-\tau)c_{XZ}s_Y(y+\alpha z)^a$};
  \draw[line width=1.5pt,->] (2.15,0.35) .. controls (3.75,1.5) .. (5.35, 0.35);
  \node at (3.75,1.5) {$\tau c_{ZY}s_Y y^a$};
  \draw[line width=1.5pt,<-] (-2.35,-0.35) .. controls (-0.75,-1.5) .. (0.85, -0.35);
  \node at (-0.25,-1.5) {$\tau c_{ZX}s_X x^a$};
  \draw[line width=1.5pt,<-] (2.15,-0.35) .. controls (3.75,-1.5) .. (5.35, -0.35);
  \node at (3.75,-1.5) {$(1-\tau)c_{YZ}s_X(x+\beta z)^a$};
  
  \draw[line width=1.5pt,->] (-3.35,-0.65) .. controls (-4.5,-2) .. (-3.35, -3.35);
  \node at (-4.5,-2) [rotate=90]{$(1-\tau)c_{XW}s_W w^a$};
  \draw[line width=1.5pt,<-] (-2.65,-0.65) .. controls (-1.5,-2) .. (-2.65, -3.35);
  \node at (-1.5,-2.5) [rotate=90]{$(1-\tau)c_{WX}s^{'}_X x^a$};
  
  \draw[line width=1.pt,|<->|] (-3,2) -- (6,2);
  \node at (1.5,2.2) {Language Transition};
  \draw[line width=1.pt,|<->|] (-5,0) -- (-5,-4);
  \node at (-5.2,-2) [rotate=90]{Ideology Transition};
  
  \end{tikzpicture}
\vspace{0.75cm}
\caption{ General scheme for ideology and language transition rates (color online).}
\label{fig:fig1}
\end{figure}
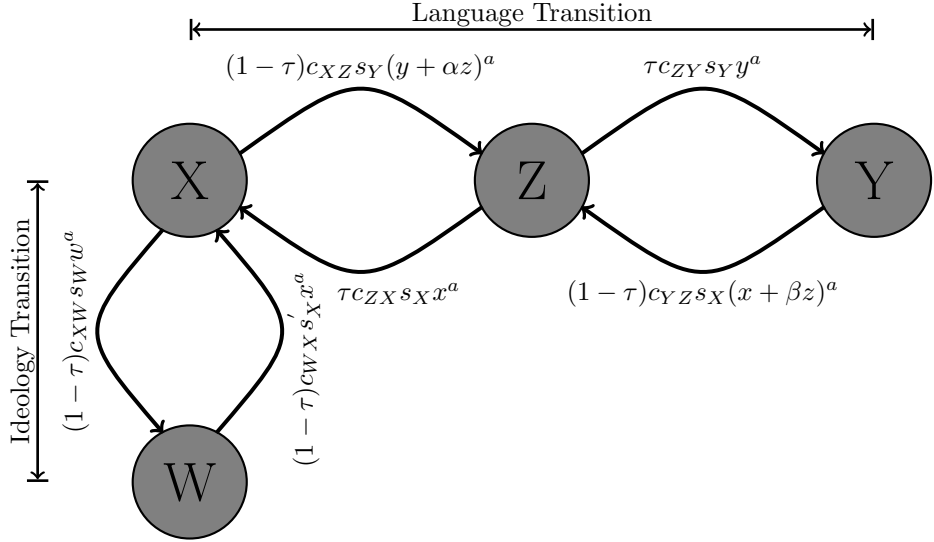
I consider a four-state model where the population living in a country can be divided into four groups denoted $X$, $Y$, $Z$ and $W$, with three types of speakers: a proportion $y\in Y$ of monolingual individuals who only speak the immigrants' low-status language, a proportion $z\in Z$ of bilingual individuals who speak both the immigrants' and the natives' language and the remainder of individuals are monolinguals who only speak the natives' high-status language, the latter being split into two groups, with a proportion $x\in X$ of individuals who support immigration and a proportion $w\in W$ which members have a nationalistic ideology and are against immigrants, where $x+y+z+w=1$. Figure \ref{fig:fig1} presents the four groups as well as the transition rates and possible transitions between groups. 

Note that by default, the model assumes that nationalism only exists among native individuals belonging to $W$, thus, individuals in $X$, $Y$ and $Z$ do not discriminate. Some interesting demographic dynamics can be contemplated in the model such as the typical three-generations pattern of complete assimilation of immigrants into the native culture found in the US and discussed in \citep{Cornell99} and \citep{Schmidt00}: a monolingual immigrant arriving in the country makes an effort to learn the native language (transition $Y\rightarrow Z$), gives birth to a child who learns both the native and the foreign language, thus still preserving the parent culture ($Z\rightarrow Z$), while that second generation individual gives birth to a child who only acquires the native language ($Z\rightarrow X$). As an adult, the latter might become nationalist ($X\rightarrow W$) or not.

With probability $\tau$ (mortality rate), adults are replaced by children and both language and ideology follow the vertical model of language transmission discussed in \citep{Minett08} and \citep{Patriarca14}, i.e., a child of a monolingual individual necessarily acquires her parent's language as her mother tongue, while a child of a bilingual individual might either acquire only one or both languages. Regarding ideology, I also assume that the first values and social norms acquired by a child are those of her parent, thus a child of an individual belonging to group $W$, initially grows up with nationalistic values while children of individuals belonging to the other groups do not discriminate anyone. Possible transitions regarding the vertical transmission mechanism are thus $X\rightarrow X$, $Y\rightarrow Y$, $Z\rightarrow Z$, $W\rightarrow W$, $Z\rightarrow X$, $Z\rightarrow Y$. As in \citep{Minett08} and \citep{Patriarca14}, there is no direct transition between groups $X$ and $Y$ because this would involve a child being unable to communicate with her parent. 

On the other hand, with probability $1-\tau$, the model follows the horizontal mechanism of language transmission in \citep{Minett08} and \citep{Patriarca14}, i.e., a bilingual adult remains bilingual over his entire life while a monolingual adult might remain monolingual or learn a second language. With regard to ideology, I assume that over life native individuals might remain nationalist or change their ideology and welcome immigrants (transitions $W\rightarrow W$ and $W\rightarrow X$). The same is valid regarding individuals in group $X$ (transitions $X\rightarrow X$ and $X\rightarrow W$). This is in line with ideology struggle in \citep{Vitanov10} and \citep{Barreira13}. Thus, possible transitions following the horizontal transmission model are $X\rightarrow X$, $Y\rightarrow Y$, $Z\rightarrow Z$, $W\rightarrow W$, $X\rightarrow Z$, $X\rightarrow W$, $W\rightarrow X$ and $Y\rightarrow Z$. The model assumes no direct transition between $W$ and $Z$ because only in the rarest case a nationalist would be interested in learning the culture and language of an immigrant. Following the most natural path, a nationalist would first change his ideology, such as in \citep{Vitanov10}, start welcoming immigrants and their culture and only then learn the language ($W\rightarrow X\rightarrow Z$).      

As can be seen in Figure \ref{fig:fig1}, transition probabilities from the bilingual group $Z$ to a monolingual group are proportional to a constant $c_{Zi}; i=\lbrace X,Y\rbrace$, to the mortality rate $\tau$, to the status of the attracting language $s_i\in\left[ 0,1\right] ;\ i=\lbrace X,Y\rbrace$ and to the size of the attracting group. As usually assumed in the literature, the relation between the status of the competing languages satisfies $s_j=1-s_i; i,j=\lbrace X,Y\rbrace; j\neq i$. On the other hand, transition probabilities in the opposite direction are proportional to a constant $c_{iZ}; i=\lbrace X,Y\rbrace$, to $1-\tau$, to the status of the attracting language and to the size of the attracting group, in this case $i+\psi z$, where in the latter either $\left( i=y\wedge\psi=\alpha\right) $ or $\left( i=x\wedge \psi=\beta\right) $. Similar to \citep{Patriarca14}, $\alpha$ and $\beta$ are the importance of bilinguals as representatives of the foreign and native languages to monolinguals in groups $X$ and $Y$, respectively. Constant parameters $c_{Zi}$ and $c_{iZ}$ reflect sociolinguistic factors \citep{Patriarca14} such as the propensity of individuals to learn a new language based on their existing linguistic skills or the provision of language resources for children \citep{Minett08}. Hence, language transition is governed by a three-state model as in \citep{Minett08} with the extension introduced in \citep{Patriarca14} such that bilinguals might influence the transition probabilities. 

Regarding the interplay between the two ideologies, the model is similar to \citep{Abrams03}, i.e., transitions occur between two-states, the nationalistic ideology and the pro-immigration one. Transition probabilities are proportional to a constant $c_i; i=\lbrace WX, XW\rbrace$, to $1-\tau$ , to the status of each ideology (either $s_X^{'}$ or $s_W=1-s_X^{'}$) and to the size of the attracting ideological group. Constants $c_{XW}$ and $c_{WX}$ play the same role as the constants in the model of \citep{Vitanov10}, i.e., larger values reflect more intense ideology conversion.


Based on the above, the evolution of the share of each group in the total population is given by the following system of non-linear ordinary differential equations:
\begin{eqnarray}
\dot{x}&=&\tau c_{ZX}s_X x^az-(1-\tau)\left[c_{XZ}(1-s_X)(y+\alpha z)^ax+c_{XW}(1-s_X^{'})w^ax-c_{WX}s_X^{'}x^aw \right]\label{x1}\\
\dot{y}&=&\tau c_{ZY}(1-s_X)y^az-(1-\tau)c_{YZ}s_X(x+\beta z)^ay\label{y1} \\
\dot{w}&=&(1-\tau)\left[c_{XW}(1-s_X^{'})w^ax-c_{WX}s_X^{'}x^aw\label{w1} \right] 
\end{eqnarray}
with $\dot{z}=-\dot{x}-\dot{y}-\dot{w}$. In this paper, in line with \citep{Minett08} and \citep{Patriarca14}, I set $a=1$, i.e., the attractiveness of a language (ideology) increases linearly with its proportion of speakers (followers). In \citep{Patriarca14}, language status, mortality rate and the parameters reflecting sociolinguistic factors are all represented by one single constant. Here, I preserve language and ideology status separately and I define a new parameter $\xi_{i}=1;\ i=\left\lbrace ZX,XZ,XW,WX,ZY,YZ\right\rbrace $ describing the interplay between mortality rate and sociolinguistic or sociocultural factors. This is equivalent to focus on the particular case in which vertical and horizontal transmission are equally likely, i.e., $\tau=0.5$, and to assume that, as in \citep{Abrams03}, all constants $c_i;\ i=\left\lbrace ZX, XZ, XW, WX, ZY, YZ\right\rbrace $ are the same, with $c_i=c=2$. Thus, taking also into account that $z=1-x-y-w$, equations (\ref{x1}) to (\ref{w1}) become: 
\begin{eqnarray}
\dot{x}&=&x\left[ s_X (1-x-y-w)-(1-s_X)(y+\alpha(1-x-y-w))+(2s_X^{'}-1)w\right]\label{x2}\\
\dot{y}&=&y\left[ (1-s_X)(1-x-y-w)-s_X(x+\beta (1-x-y-w))\right] \label{y2} \\
\dot{w}&=&xw(1-2s_X^{'})\label{w2}
\end{eqnarray}
The phase space is given by the unit tetrahedron $\Omega=\left\lbrace \theta\in[0,1]^3:x+y+w\leq 1 \right\rbrace$ and the system above has the following isolated fixed points: $(1,0,0)$ and $(\bar{x},\bar{y},0)$, where $\bar{x}=\frac{(1-s_X)\left[ 1-s_X(1+\beta)\right] }{1-s_X+s_X^2-s_X(1-s_X)(\alpha+\beta)}$ and $\bar{y}=\frac{s_X\left[s_X-\alpha(1-s_X)\right] }{1-s_X+s_X^2-s_X(1-s_X)(\alpha+\beta)}$, the latter fixed point belongs to $\Omega$ if $\frac{\alpha}{1+\alpha}<s_X<\frac{1}{1+\beta}$. The following sets of fixed points are also in $\Omega$: $(0,0,w)$ and $(0,y,1-y)$. The stability of the isolated fixed points can be determined by the analysis of the eigenvalues of the linearized system, while I use a qualitative analysis of the evolution of the trajectories to study stability of the two sets of fixed points. The eigenvalues of the Jacobian matrix evaluated at $(1,0,0)$ are $\lambda_i=\left\lbrace -s_X+(1-s_X)\alpha;-s_X;1-2s_X^{'}\right\rbrace$, which are all negative provided $s_X^{'}>0.5$ and $s_X>\frac{\alpha}{1+\alpha}$, i.e., a population of monolinguals of the native language is an asymptotically stable state provided the status of the native language is above a given threshold and the status of the pro-immigration (assimilationist) ideology is stronger than the nationalistic one. On the other hand, the eigenvalues of the Jacobian matrix evaluated at $(\bar{x},\bar{y},0)$ have mixed signs
, thus the fixed point $(\bar{x},\bar{y},0)$ is unstable. 

\textbf{Theorem 1:} when the status of the native language is within an intermediate level given by $\underline{s_X}=\frac{\alpha}{1+\alpha}<s_X<\frac{1}{1+\beta}=\overline{s_X}$ and the status of the pro-immigration ideology is above that of the nationalistic ideology, $s_X^{'}>0.5$, the population of the country either evolves to a monolingual group of native language speakers, i.e., immigrants get assimilated and nationalism disappears, or to a bilingualism state composed of two segregated groups, one of nationalistic individuals, all speaking the native language, and the other of monolinguals speaking the foreign language. In such a context, in contrast with the widespread literature, monolingualism can actually be a good state and bilingualism an undesired one.

\textbf{Proof:} when $(\underline{s_X}<s_X<\overline{s_X})\wedge \left( s_X^{'}>0.5\right) $, any trajectory starting in the interior of $\Omega$ satisfies $\dot{w}<0$. Moreover, $\dot{x}=0$ at any point located on the plane $\mathcal{P}_x:\ x=(1-y-w)-\frac{(1-s_X)y-(2s_X^{'}-1)w}{s_X-\alpha(1-s_X)}$ and $\dot{y}=0$ at any point located on the plane $\mathcal{P}_y:\ x=(1-y-w)\left( 1-\frac{s_X}{1-s_X\beta}\right) $. At the boundary of $\Omega$ given by $w=0$, $P_x$ and $P_y$ are lines intercepting each other at the fixed point $(\bar{x},\bar{y},0)$. $\mathcal{P}_x$ crosses the boundaries of $\Omega$ at the vertex $(1,0,0)$ and at $\left( 0,\frac{s_X-\alpha(1-s_X)}{1-\alpha(1-s_X)},0\right) $ and $\left( 0,\frac{2s_X^{'}-1}{2s_X^{'}-s_X},\frac{1-s_X}{2s_X^{'}-s_X}\right)  $. Any trajectory starting in the interior of $\Omega$ and above (below) $P_x$ satisfies $\dot{x}<0$ ($\dot{x}>0$), hence in the invariant manifold $(x,0,0)$, $\dot{x}$ is necessarily non-negative. On the other hand, $\mathcal{P}_y$ crosses the boundaries of $\Omega$ at the vertices $(0,1,0)$, $(0,0,1)$ and at $\left( 1-\frac{s_X}{1-s_X\beta},0,0\right) $. Any trajectory starting in the interior of $\Omega$ and above (below) $P_y$ satisfies $\dot{y}<0$ ($\dot{y}>0$), hence in the invariant manifold $(0,y,0)$, $\dot{y}$ is necessarily non-negative. Consequently, any population with initial conditions located both below $P_x$ and above $P_y$ satisfies $\dot{x}>0$; $\dot{y}<0$ and $\dot{w}<0$, thus converging in the long run to the asymptotically stable state $(1,0,0)$, becoming a monolingual population of speakers of the native language, in which nationalism becomes extinct and immigrants become linguistically and culturally assimilated if one follows \citep{Shafir95} and adopts language as a proxy for culture. On the other hand, any initial condition located both above $P_x$ and below $P_y$ satisfies $\dot{x}<0$; $\dot{y}>0$ and $\dot{w}<0$, thus converging to a neutrally stable state in the invariant manifold $(0,y,1-y)$, i.e., a segregated bilingual society with immigrants speaking the foreign language and nationalistic individuals speaking the native language, i.e., both groups $Y$ and $W$ survive. Any other interior initial condition necessarily converges to either $(1,0,0)\vee(0,y,1-y)$ $\blacksquare$
\begin{figure}[htp]
\centering
\begin{tabular}{cc}
\epsfig{file=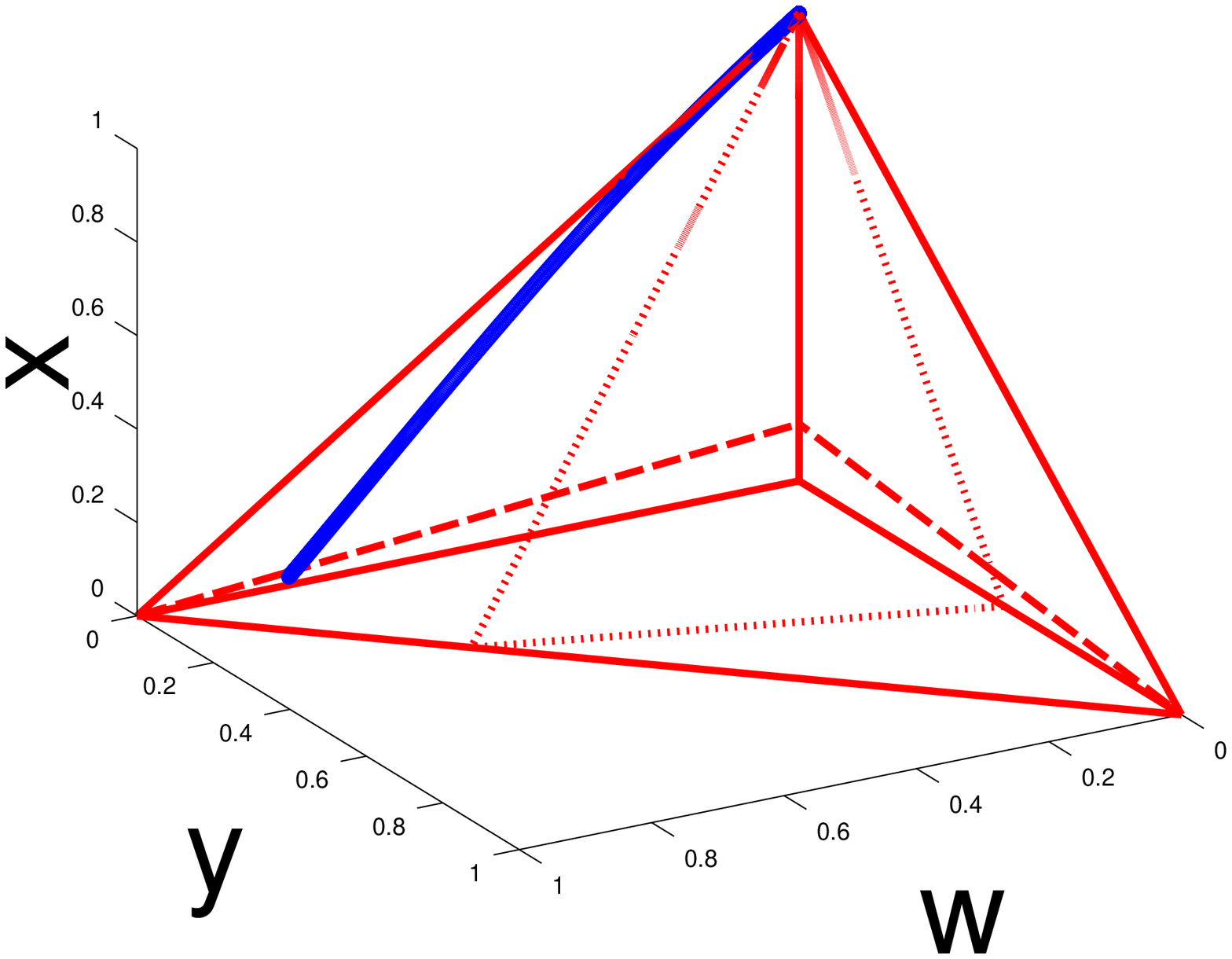,height=7cm,width=7cm,angle=0}&\epsfig{file=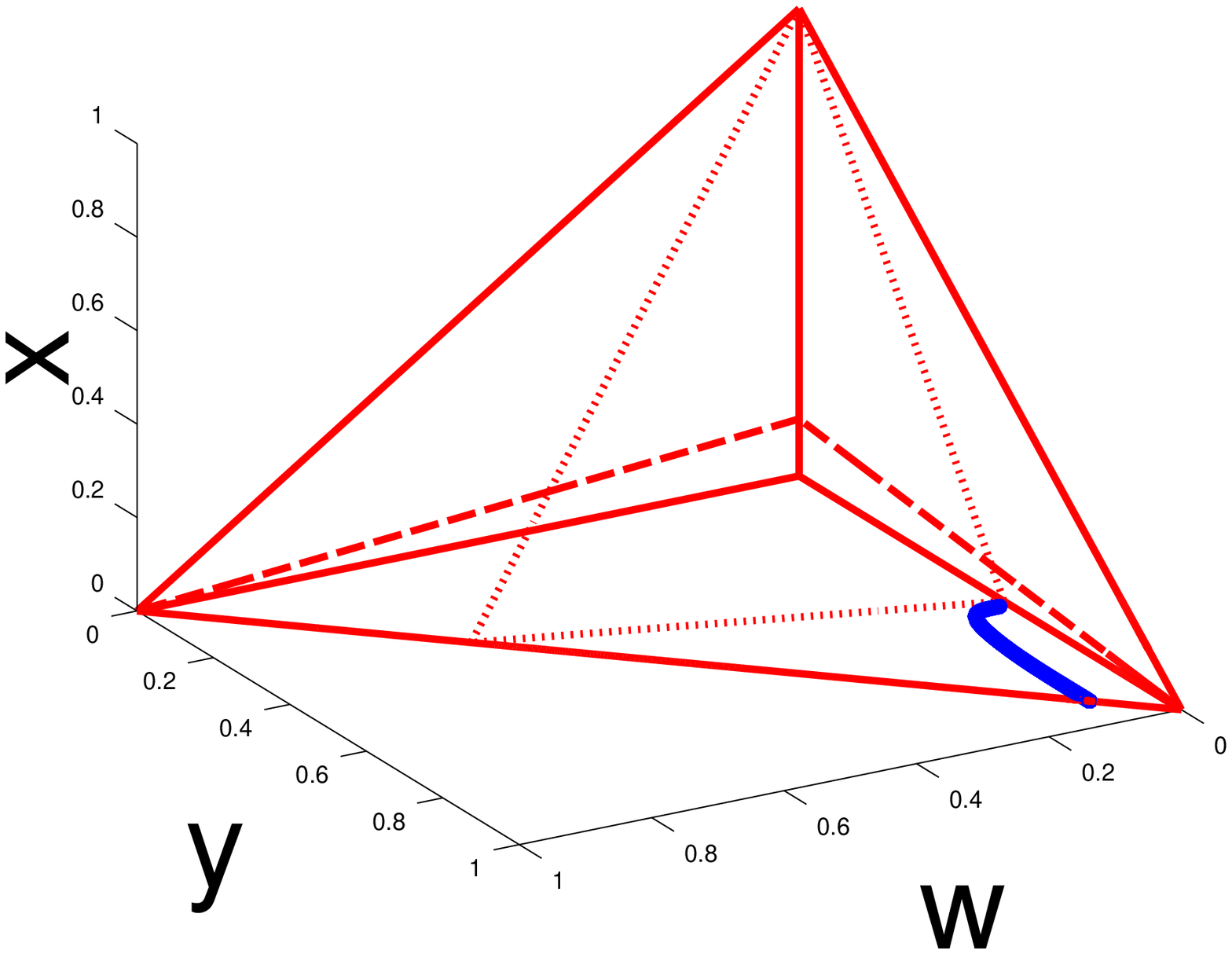,height=7cm,
width=7cm,angle=0}\\
\end{tabular}
\caption{ Phase spaces for parameters: $s_X=0.575$; $s_X^{'}=0.6$; $\alpha=0.2$; $\beta=0.6$; initial conditions $(0.05,0.05,0.8)$ (left panel) and $(0.1,0.7,0.1)$ (right panel); dotted (dashed) lines represent plane $\dot{x}=0$ ($\dot{y}=0$); step size $\Delta t=0.05$ (color online).}
\label{fig:fig2}
\end{figure} 

\begin{figure}[htp]
\centering
\begin{tabular}{cc}
\epsfig{file=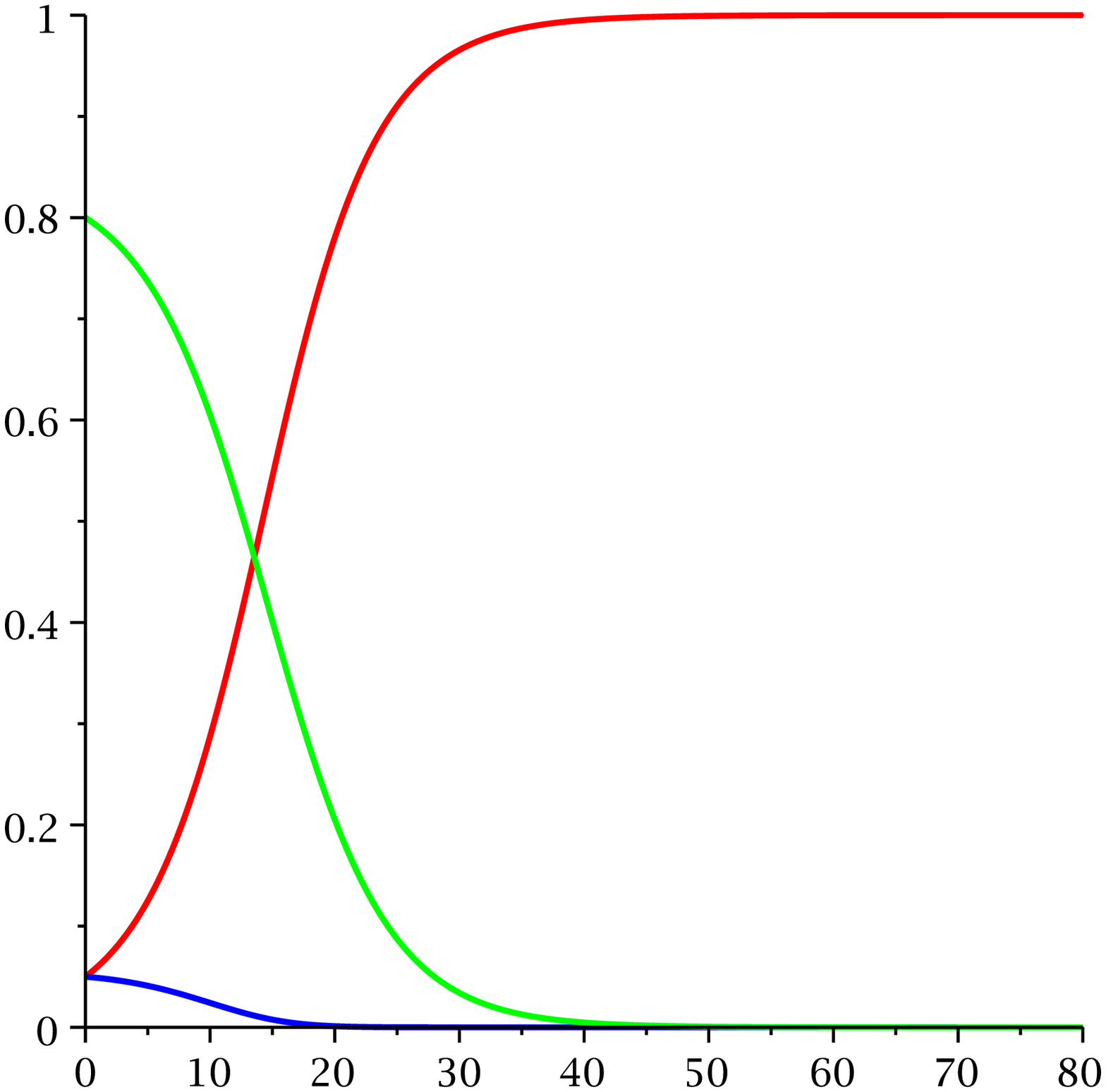,height=4.5cm,width=7cm,angle=0}&\epsfig{file=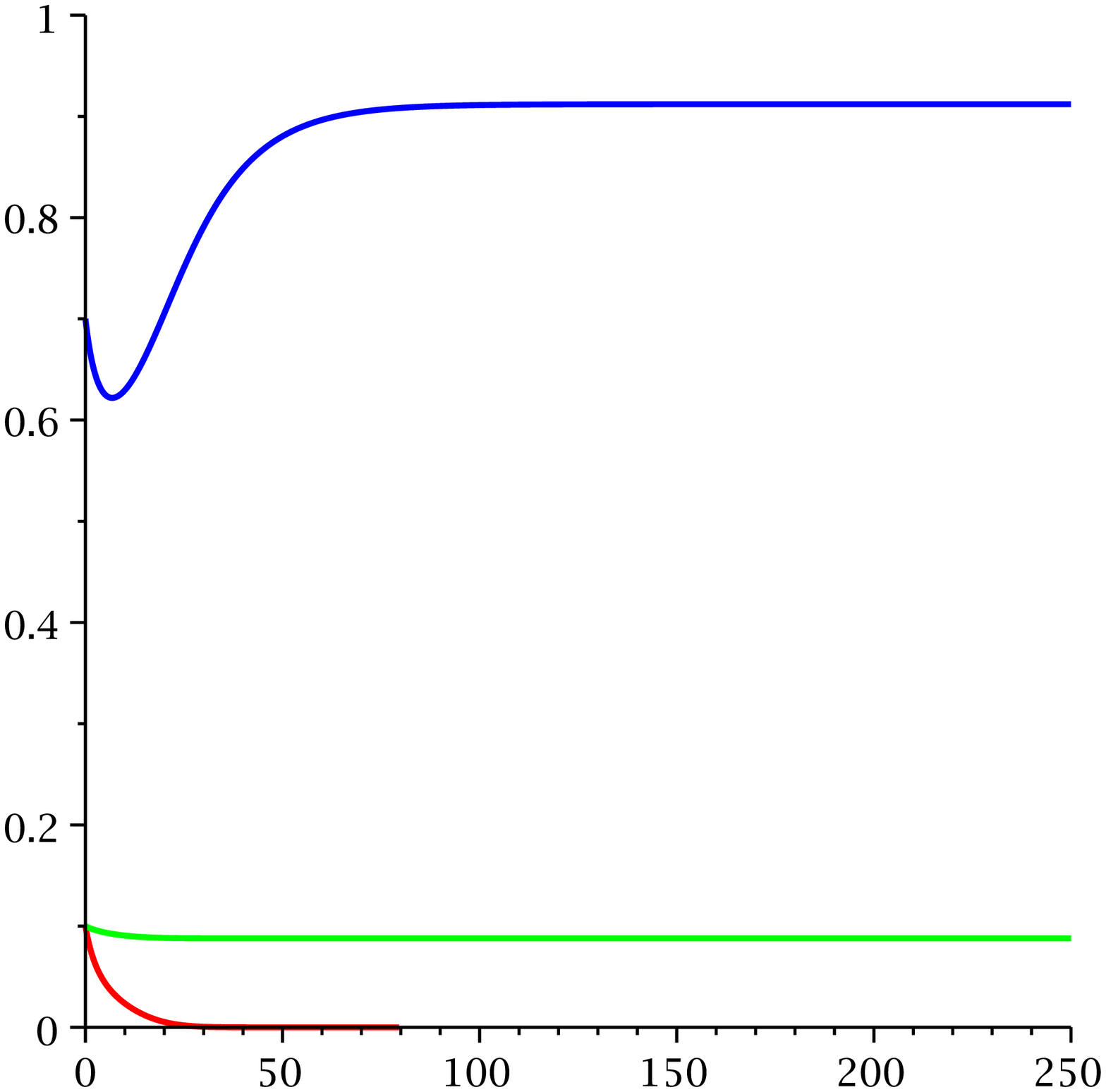,height=4.5cm,
width=7cm,angle=0}\\
\end{tabular}
\vspace{0.75cm}
\caption{ Time evolution for parameters: $s_X=0.575$; $s_X^{'}=0.6$; $\alpha=0.2$; $\beta=0.6$; initial conditions $(0.05,0.05,0.8)$ (left panel) and $(0.1,0.7,0.1)$ (right panel); step size $\Delta t=0.05$; red: pro-immigration natives; green: nationalists; blue: foreign language monolinguals (color online).}
\label{fig:fig3}
\end{figure}

Figures \ref{fig:fig2} and \ref{fig:fig3} present respectively the phase spaces and the time evolutions for two simulations using the following parameters: $s_X=0.575$; $s_X^{'}=0.6$; $\alpha=0.2$; $\beta=0.6$, i.e., a country in which the pro-immigration ideology status is one and a half times stronger than the nationalistic ideology, the native language has a stronger status than the foreign language and bilinguals are moderately seen as representatives of the native language for the monolinguals in $Y$ but natives in $X$ weekly see bilinguals as representatives of the immigrants' language. Initial conditions are $(x_0,y_0,w_0)=(0.05,0.05,0.8)$ (left panels) and $(x_0,y_0,w_0)=(0.1,0.7,0.1)$ (right panels). With the former initial conditions, the share of group $X$ is monotonically increasing while nationalists shift their ideology and over time monolingual immigrants become bilinguals further becoming fully assimilated, i.e., $Y\rightarrow Z\rightarrow X$. On the other hand, with the latter initial conditions, one can see the evolution towards a segregated country with two monolingual groups and nationalism. As in the first simulation, nationalism is monotonically decreasing, but once pro-immigrants in $X$ disappear, the share of nationalists that remains is no more influenced by those to shift their ideology and monolinguals of the native language remain in an isolated nationalistic enclave, without contact with monolinguals of the foreign language.

\textbf{Theorem 2:} in the absence of nationalism, bilingualism is not stable and one language eventually takes over the entire population as in \citep{Abrams03} and \citep{Minett08}.

\textbf{Proof:} when $w=0$, the phase space becomes the unit triangle $\Omega=\left\lbrace \theta\in[0,1]^2:x+y\leq 1 \right\rbrace$. The eigenvalues of the Jacobian matrix evaluated at point $(0,0)$, i.e., $z=1$, are $\lambda_i=\left\lbrace s_X-\alpha(1-s_X);1-s_X-s_X\beta\right\rbrace$ which are both negative if $\alpha>\frac{s_X}{1-s_X}\wedge\beta>\frac{1-s_X}{s_X}$. Such a condition can never be satisfied given $\alpha,\beta\in[0,1]$ by definition. Thus, when bilinguals do affect the transition probabilities as in \citep{Patriarca14}, bilingualism stability depends on the way the conversion rate constants $k_{ZX}$, $k_{XZ}$, $k_{ZY}$, $k_{YZ}$ are defined in equations (\ref{P1}) and (\ref{P2}). If language status is to be explicitly defined in the system equations, bilingualism is unstable $\blacksquare$

\section*{Acknowledgements}
I acknowledge the research support ``Programa de Incentivo \`a Produtividade em Ensino
e Pesquisa'' from PUC-Rio.

\end{document}